\documentclass[aps,prb,amsmath,amssymb,reprint]{revtex4-1}
\usepackage{chemfig}
\usepackage{amsmath}
\usepackage{amssymb}
\usepackage{color}
\usepackage{times}
\usepackage{epsf}
\usepackage{graphicx}
\usepackage{epsfig}
\usepackage{epstopdf}
\usepackage{dcolumn}
\usepackage{bm}
\usepackage{float}
\usepackage{url}
\usepackage{CJK}
\usepackage{ulem}
\usepackage{makecell}

\def\<{\langle}
\def\>{\rangle}
\def\({\left(}
\def\){\right)}
\def\[{\left[}
\def\]{\right]}

\def\sin{\mathop{\mathrm{sin}}\nolimits}
\def\cos{\mathop{\mathrm{cos}}\nolimits}

\begin{document}

\title{Theoretical prediction of interfacial
thermal conductance at high temperature across solid-solid interfaces}

\author{Jinxin Zhong}
\affiliation{Center for Phononics and Thermal Energy Science, China-EU Joint Lab for Nanophononics,
School of Physics Science and Engineering, Tongji University, 
Shanghai 200092, China}

\author{Zhiguo Wang}
\affiliation{School of Physics Science and Engineering, Tongji University, 
Shanghai 200092, China}

\author{Xiaobo Li}
\affiliation{Huazhong University of Science and Technology, Wuhan, China}

\author{Jun Liu}
\affiliation{Department of Mechanical and Aerospace Engineering, North Carolina State University, Raleigh, NC 27695, USA}

\author{Jun Zhou}
\email{zhoujunzhou@tongji.edu.cn}
\affiliation{NNU-SULI Thermal Energy Research Center (NSTER) and Center for
  Quantum Transport and Thermal Energy Science (CQTES), School of Physics and Technology, Nanjing Normal
  University, Nanjing 210023, China}

\date{\today}

\begin{abstract}
	The existed theories and methods for calculating interfacial thermal conductance of solid-solid interface lead to diverse values that deviate from experimental measurements. In this letter, We propose a model to estimate the ITC at high temperature without comprehensive calculations, where the interface between two dissimilar solids can be treated as an amorphous thin layer and the coordination number density across interface becomes a key parameter. Our model predicts that the ITCs of various interfaces at 300K are in a narrow range: 10$^{7}$\,W·m$^{-2}$·K$^{-1}$\,$\sim $10$^{9}$\,W·m$^{-2}$·K$^{-1}$, which is in good agreement with the experimental measurement.
\end{abstract}

\maketitle

A temperature jump at interface between two dissimilar materials is
inevitable when a heat current flows across the interface. The
origin is the non-zero interfacial thermal resistance (ITR).
\cite{swartz} Many powerful theories and theoretical tools have been developed to
calculate ITR, for example, acoustic
mismatch model,\cite{khalatnikov} diffuse mismatch
model,\cite{swartz}, molecular dynamics (MD),\cite{maiti} lattice
dynamics (LD),\cite{young,pettersson} and atomic Green's 
function (AGF).\cite{zhang} In these theories and tools, material properties
such as phonon density of states, sound velocity, and Debye cut-off frequency are
required. The detailed atomic structures are also required in the MD,
LD, and AGF simulations. Consequently, the calculated values of ITR highly
depends on materials and different methods lead to very different results.

The experimental measured interfacial thermal conductance (ITC), which
is the inversion of ITR, of well treated clean interfaces at high temperature
are in a narrow range: 10$^{7}$\,W·m$^{-2}$·K$^{-1}$\,$\sim 8\times$10$^{8}$\,W·m$^{-2}$·K$^{-1}$.\cite{wilson} There
seems to be a universal mechanism which limits the maximum value of
ITC. Stoner and Maris \cite{stoner} found that the ITC of
Ti/Sapphire and Al/Sapphire is only two times larger than
the ITC of Pb/Sapphire. This contradicts the calculation results which
show that the ITCs of Ti/Sapphire and Al/Sapphire are two
orders of magnitude larger than the ITC of Pb/Sapphire. Hohensee
{\sl et al.}\cite{hohensee} eliminated the influence from the weak
bonding by applying high pressure. They found that the ITC of metal/diamond
interface under high pressure weakly depends on the type of metal. Lu {\sl
  et al.} \cite{Lu} pointed out that this is possibly attributed to
the contribution from heat transfer channel of electron-phonon
coupling. The problem is still unsolved because the measured ITCs of GaN/AlN,
TiN/MgO, and SrRuO$_3$/SrTiO$_3$ are 6.2, 7.2, and 8.0$\times$10$^{8}$\,W·m$^{-2}$·K$^{-1}$, respectively.\cite{Koh,Costescu,wilson} These systems have very
similar ITC although there is no conduction electron nor
electron-phonon coupling channel. Moreover, it has been observed that the ITC at the interface formed by amorphous materials is higher than that of crystalline materials.\cite{Giri} The thermal conductance at the interface of a-SiO$_{2}$/a-Al$_{2}$O$_{3}$ is as high as 7.1$\times$10$^{8}$\,W·m$^{-2}$·K$^{-1}$.\cite{Fong} Wilson {\sl et al.}\cite{wilson}
have found a clue which stimulates us to look into this problem. They
pointed out that the maximum value of ITC is linearly correlated to
the product of Debye velocity $v_{\rm D}$ and volumetric heat capacity
$C_{\rm v}$. The values of $v_{\rm D}C_{\rm v}$ of different materials are on the
order of $10^{10}$\,W·m$^{-2}$·K$^{-1}$. Giri {\sl et al.} \cite{Giri} have found another aspect of this problem. They found that a better match of elastic moduli between crystalline materials forming the interface usually results in a higher ITC. Another interesting phenomenon is equally noteworthy. The thermal conductivity of amorphous materials are also in a narrow range: about 10$^{-1}$\,W·m$^{-1}$·K$^{-1}$. Dividing this magnitude by the magnitude of the interface width (10$^{-10}$ m) can get 10$^{9}$\,W·m$^{-2}$·K$^{-1}$, which equals to the magnitude of ITC. Baesd on these facts, we postulate that the interface region could be regarded as an amorphous or disordered system.

\begin{figure}[htp]
	\centering
	\includegraphics[width=0.8\linewidth]{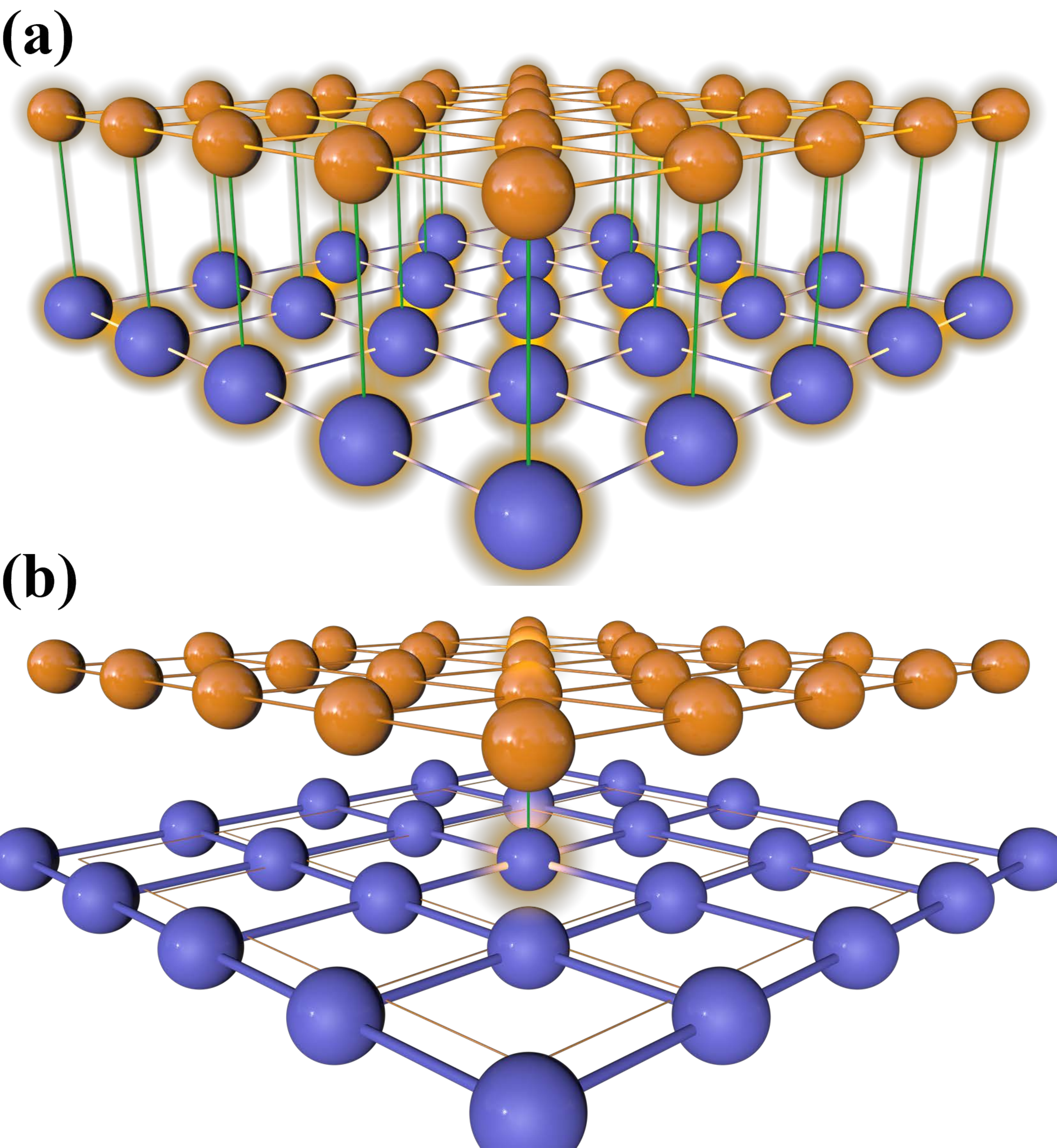}
	\caption{(Color online) Schematic diagram of perfect (a) and dissimilar (b) interface. Balls represents atoms and lines represents the bonds between them.} 
	\label{fig1}
\end{figure}

In this letter, we propose a method to estimate the ITC at high
temperature without comprehensive calculations. We point out that the 
interface between two dissimilar solids can be treated as an
amorphous thin layer in which the coordination number density across interface is the
key parameter. The recently developed thermal transport theory
of amorphous solids by us are extended to calculate the ITC.

The ITC can be written as: \cite{Xi}
\begin{equation}
h=\frac{1}{R}=\alpha n_{\text b} g,
\label{ITR}
\end{equation}
where $\alpha$ is a dimensionless factor corresponding to the energy transfer efficiency between atoms at the interface. $n_{\text b}$
is the two-dimensional number density of coordination number across the interface. $g=C v_{\rm eff}
/\delta$ is the heat transfer between two atoms per unit time per
temperature drop where $C$ is the per particle heat capacity,
$v_{\rm eff}$ is the effective velocity in the thin layer, and $\delta$ is
the average distance between neighboring atoms. 

\begin{figure}[htp]
	\centering
	\includegraphics[width=1 \linewidth]{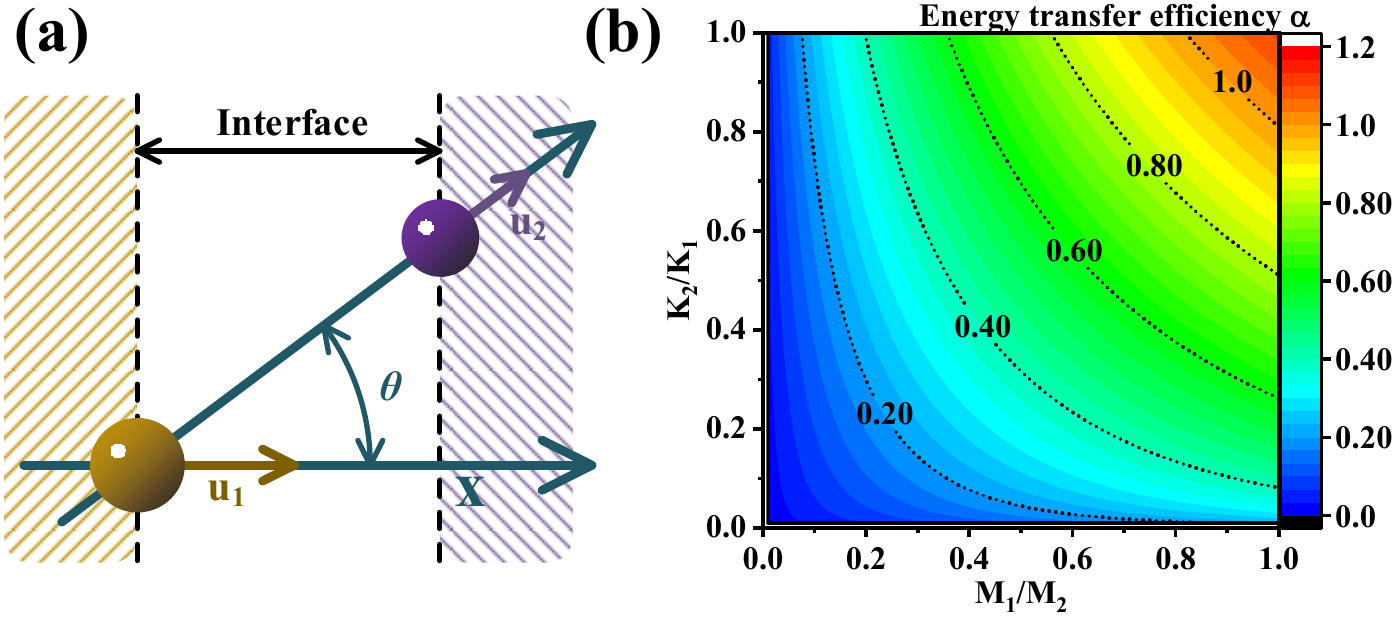}
	\caption{(Color online) (a) Illustration of the interaction between two atoms at the interface. (b) Calculated energy transfer efficiency as a variation of $K_2/K_1$ and $M_1/M_2$, with $K_{\text 3}=(K_{\text 1}+K_{\text 2})/2$.} 
	\label{fig2}
\end{figure}

We now analyze the parameters in Eq.\,(\ref{ITR}). 
	Energy transfer efficiency is considered to be related to the matching degree of elastic modulus\cite{Giri} and atomic mass at the interface.
	A qualitatively explanation of $\alpha$ is given here by calculating the energy transfer efficiency between two uncorrelated harmonic oscillators following Einstein's approach\cite{Kaviany}. 
	Considering a set of two atoms, as shown in Fig.\,\ref{fig2}(a), 
	a dislplacement of magnitude $u_{\text 1}$ of the left atom in the x direction will cause the right atom moving along the line between them by an amount $u_{\text 2}$. 
	The angle formed by these two diretions is $\theta$. 
	The change in the distance between them is then $(u_{\text 2}-u_{\text 1}\cos\theta)$, and the equation of motion for the left atom in the x direction is $M_{\text 1}\dfrac{d^{2}x}{dt^{2}} =K_{\text 3}(u_{\text 2}-u_{\text 1}\cos\theta)\cos\theta$, where $K_{\text 3}$ is the interface spring constant and $M_{\text 1}$ is the mass of the left atom.
	Assuming that the atomic motions are sinusoidal and uncorrelated, thus $u_{\text 1}=A_{\text 1}\sin(\omega_{\text 1} t)$ and $u_{\text 2}=A_{\text 2}\sin(\omega_{\text 2} t+\phi_{\text i})$, with $A_{\text 1,2}=\sqrt{2\hbar \omega_{\text 1,2}/K_{\text 1,2}}$ and $\omega_{\text 1,2}=\sqrt{K_{\text 1,2}/M_{\text 1,2}}$, where  $\phi_{\text i}$ is a random phase angle, $K_{\text 1,2}$ and $M_{\text 1,2}$ are force constant and atomic mass, respectively.
	Considering a time of one-half of the period of oscillation $(\pi /\omega_{\text 1})$, we can get the energy change $(\overline{(dE)^{\text 2}})^{1/2}=\dfrac{\text 1}{\text 2}K_{\text 3}A_{\text 1}A_{\text 2}\dfrac{\omega_{\text 1}\omega_{\text 2}}{\omega_{\text 1}^{\text 2}-\omega_{\text 2}^{\text 2}}(1+\cos(\dfrac{\omega_{\text 1}}{\omega_{\text 2}} \pi))^{1/2}$, and the energy transfer effciency is then defined by $(\overline{(dE)^{\text 2}})^{1/2}/(\hbar \omega_{\text 1})$. 
	Fig.\,\ref{fig2}(b) shows the calculated energy transfer efficiency as a function of $K_2/K_1$ and $M_1/M_2$. 
	It is concluded that the energy transfer efficiency $\alpha$ is positively correlated with $K_2/K_1$, which is in good agreement with Giri' work\cite{Giri}.

$n_{\text b}$ depends on atomic structures and crystalline orientations of material 1 and material 2. The $n_{\text b,c}$ at interface of crystalline materials can be calculated by $n_{\text b,c}=\beta n_{\rm p}$.
For example, if the structure of material 1 is fcc and the
(100) plane is parallel to the interface, then the number density of
plane 1 ($n_{\rm p}$) is ${\text 2}/a_{\rm p}^{2}$ where $a_{\rm p}$ is the lattice
constant. $\beta$ is the lattice matching, also can call that the bonding rate, and it's only chosen as 1 for the perfect interface which is shown in Fig.\,\ref{fig1}(a). 
However, in realistic cases, the lattice matching can be far smaller than 1 due to the dissimilar of atomic structures and crystalline orientations of material 1 and material 2. The determination of $\beta$ requires numerical simulation.
In order to evaluate the lattice matching at the interface between two dissimilar crystalline solids, the interface is treated as an amorphous thin layer. 
The amorphous thin layer is composed of two atom layers belonging to material 1 and material 2, respectively.
The atom arrangement of each layer is determined by the chosen crystallographic plane of the given material.
In our simulation, the distance between the top and bottom layer is set as the summation of their Van Der Waals Radius ($\delta$). The atom group of the bottom layer is set large enough to be considered as infinite, while the size of top layer is finite for the sake of counting the lattice matching. 
The atom group of top layer is randomly translated and rotated in the range of the Wigner-Seitz cell of the given bottom layer, and the moving vector and rotating angle are generated through a Monte Carlo simulation.
Once the atom group of top layer is translated or rotated, the distance between atoms of the top layer and bottom layer will be calculated. 
If the distance between atoms is smaller than 1.01$\delta$ (estimated by Lindemann Criterion of Melting), it will be considered as matching.
After 10W times translating and rotating, the simulation is convergent and the lattice matching is obtained. 

And the determination of $n_{\text b,a}$ at amorphous materials interface also requires numerical simulation. 
In order to evaluate the lattice matching at the interface between two dissimilar amorphous solids, the structure of amorphous solids are simplified into random close-packing sphere.
The positions of the atomic sphere are random, thus we adopted the Monte Carlo method to generate an equivalent amorphous structure. 
the number  of atoms ($N_{as}$) to be generated in a given box is $N_{as}=L^{3}\times \rho$, where $L$ is the size of the simulation box and $\rho$ is the measured number density. 
In our simulation, there are two cubic boxes aligning in one direction, each with $L=10$\,nm. The distance between the two boxes is set as $\delta$. 
Then the position of particles is randomly generated and recorded. 
If the distance between the atoms belonging to different boxes is smaller than 1.01$\delta$, it will be considered as matching at the interface.
After all atoms have been generated, the total matching number $N_{matching}$ is obtained. Then $n_{\text b,a}$ is evaluated from $n_{\text b,a}=N_{matching}/L^{2}$. 
After 1W times generating, the simulation converges and the final $n_{\text b,a}$ is obtained.

The average per particle heat capacity at high temperature is $3k_{\rm B}$.
The effective sound velocity can be written as $v_{\rm eff}=\left(\sum_{1,j}v_{1,j}+\sum_{2,j}v_{2,j}\right)/6$. 
The inter-atomic distance is chosen as $\delta\approx (d_{1}+d_{2})$ where $d$ is the Van Der Waals Radius.

\begin{figure}[htp]
	\centering
	\includegraphics[width=0.9\linewidth]{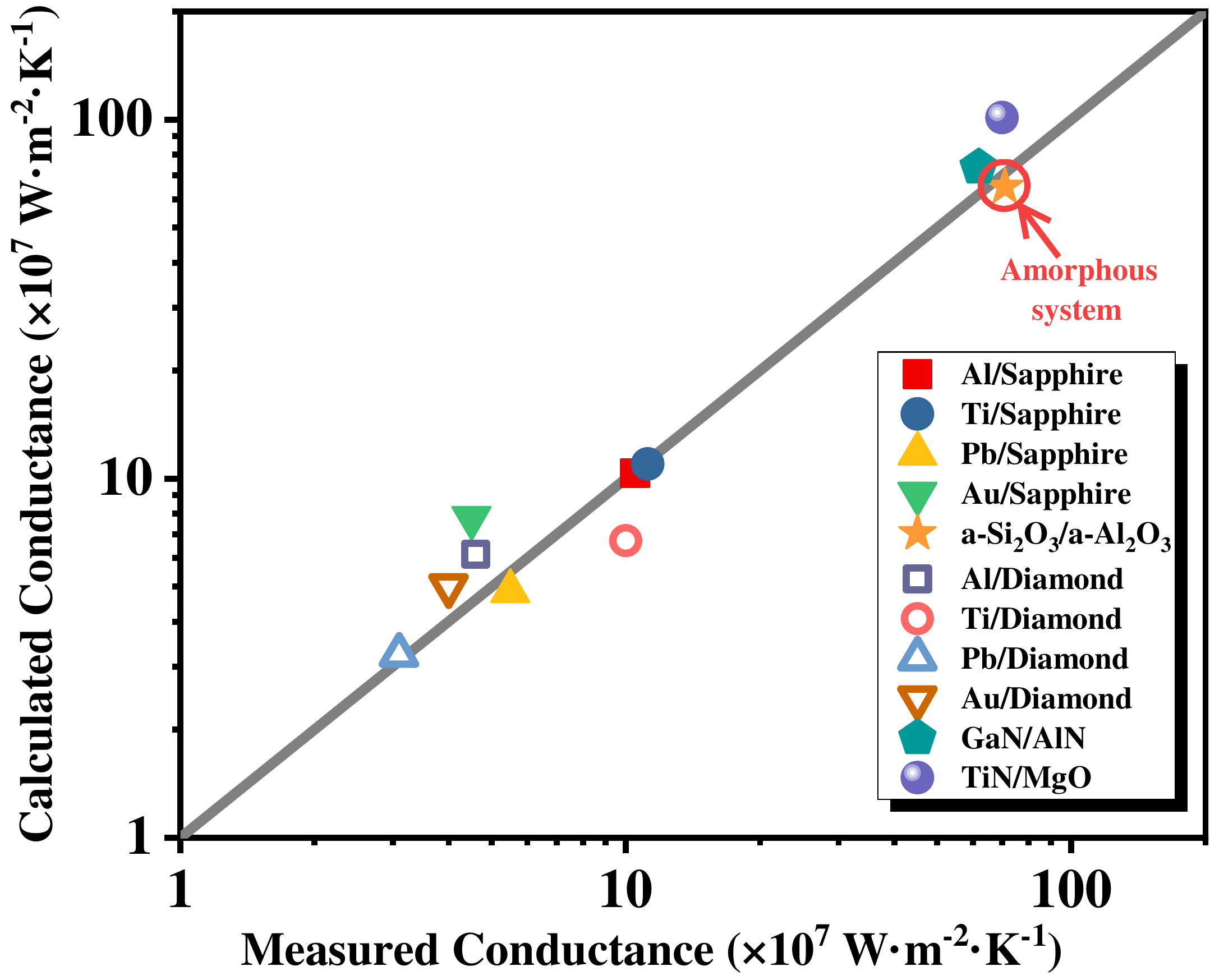}
	\caption{(Color online) Calculated thermal conductance of eleven different interfaces at 300 K vs the measured values. Measured date for thermal conductance of Al, Ti, Pb, Au/Sapphire(2$\bar{1}\bar{1}$0) and Al, Ti, Pb, Au/Diamond(100) are from Ref.\citenum{stoner}, a-SiO$_{2}$/a-Al$_{2}$O$_{3}$ from Ref.\citenum{Fong}, GaN/AlN from Ref.\citenum{Koh}, TiN(111)/MgO(111) from Ref.\citenum{Costescu}, SrRuO$_{3}$(001)/SrRuO$_{3}$(001) from Ref.\citenum{wilson}.}
	\label{fig3}
\end{figure}

We now exemplify our model by calculating the ITR of Al/Sapphire,
Ti/Sapphire, and Pb/Sapphire interfaces. 

For Al, fcc(100): $a=4.05\,{\text \AA}$, $n=2/a^{2}=1.22\times
10^{19}$\,m$^{-2}$, $v_{\rm L}=6500$\,m·s$^{-1}$, $v_{\rm T}=3300$\,m·s$^{-1}$. $d=2.10\,{\text \AA}$.
For Ti, hcp(0001): $a=2.94\,{\text \AA}$, $n=1/a^{2}=1.34\times
10^{19}$\,m$^{-2}$, $v_{\rm L}=6100$\,m·s$^{-1}$, $v_{\rm T}=3150$\,
m·s$^{-1}$. $d=2.15\,{\text \AA}$.
For Pb, fcc(100): $a=5.06\,{\text \AA}$, $n=2/a^{2}=7.83\times
10^{18}$\,m$^{-2}$, $v_{\rm L}=2160$\,m·s$^{-1}$, $v_{\rm T}=705$\,
m·s$^{-1}$. $d=2.05\,{\text \AA}$. For Al$_2$O$_3$, O-plane: $a=4.76\,{\text \AA}$,
$n=3/(\sqrt{3}a^{2}/2)=1.53\times 10^{19}$\,m$^{-2}$ and $d=1.55\, {\text \AA}$; Al-plane: $n=1/(\sqrt{3}a^{2}/2)=5.01\times 10^{18}$\,m$^{-2}$ and $d=2.10\,{\text \AA}$; $v_{\rm L}=11000$\, m·s$^{-1}$, $v_{\rm T}=6450$\,m·s$^{-1}$.

According to the simulation, $n_{\rm b}=6.85\times 10^{17}$\,m$^{-2}$, $6.70\times 10^{17}$\,m$^{-2}$, $4.29\times 10^{17}$\,m$^{-2}$, $1.55\times 10^{18}$\,m$^{-2}$, $1.75\times 10^{18}$\,m$^{-2}$, and $9.69\times 10^{17}$\,m$^{-2}$ for Al/Al-plane, Ti/Al-plane, Pb/Al-plane, Al/O-plane, Ti/O-plane and Pb/O-plane,  respectively. 

The interfacial thermal conductance is evaluated through Eq.\,(\ref{ITR}) after averaging according to the atomic number ratio of compounds. It is obtained that $h=1.02\times 10^{8}$\,W·m$^{-2}$·K$^{-1}$, $1.09\times 10^{8}$\,W·m$^{-2}$·K$^{-1}$, and $0.48\times 10^{8}$\,W·m$^{-2}$·K$^{-1}$ for Al/Sapphire, Ti/Sapphire, and Pb/Sapphire, respectively. It shows that the ITCs of Al/Sapphire and Ti/Sapphire are two times larger than the ITC of Pb/Sapphire, which is in good agreement with Stoner and Maris' work\cite{stoner}. 

We further estimated the ITCs of (Al, Ti, Pb, Au)/diamond, GaN/AlN, TiN/MgO, and SrRuO$_{3}$/SrTiO$_{3}$ at 300K with parameters given in Table.\,\ref{table1}, and the comparison between calculated and measured conductance is shown in Fig.\,\ref{fig3}.
These results indicate that the calculated ITCs at high temperature are in a range: 10$^{7}$\,W·m$^{-2}$·K$^{-1}$\,$\sim $10$^{9}$\,W·m$^{-2}$·K$^{-1}$, which are consistent with the measured conductance in magnitude.\cite{wilson} 
And the high ITC at the amorphous interface of a-SiO$_{2}$/a-Al$_{2}$O$_{3}$ is originated from the larger coordination number density $n_{\text b}$.

In conclusion, we have proposed a model to estimate the ITC at high temperature without comprehensive calculations. 
We point out that the interface between two dissimilar solids can be treated as an amorphous thin layer in which the coordination number density across interface is the key parameter.
The method could reproduce the observation that the ITCs of Al/Sapphire and Ti/Sapphire are two times larger than the ITC of Pb/Sapphire. 
And further calculation on more interfaces shows the ITCs at 300K are in a narrow range: 10$^{7}$\,W·m$^{-2}$·K$^{-1}$\,$\sim $10$^{9}$\,W·m$^{-2}$·K$^{-1}$, which is in good agreement with the experimental measurement. 
The high ITC at the interface of the amorphous materials is originated from the larger coordination number density.
\\
\\
This work was supported by National Key R{\&}D Program of China
(No. 2017YFB0406004), National Natural Science Foundation of China
(No. 11890703).

\begin{table*}[bp]
	\caption{Energy transfer efficiency, effective velocity, and average distance \cite{Batsanov} of typical interfaces used in the calculations. The number density of coordination number across the interface is calculated through Monte Carlo simulation. The interface thermal conductance is evaluated by Eq.\,(\ref{ITR}) after averaging according to the atomic number ratio of compounds.}
	\begin{tabular}{c c c c c c c}
		\hline
		Interface & $\alpha$ & {$v_{\rm eff}$ [m·s$^{-1}$]} & Plane/Plane & $\delta$ [$\rm{\AA}$] & $n_{\text b}$ [$\times$10$^{18}$\,m$^{-2}$] & $h$ [$\times$10$^{8}$\,W·m$^{-2}$·K$^{-1}$
		] \\
		\hline
		Al/Sapphire & 0.125 & 6166.67 & \makecell{Al(100)/Al(0001) \\ Al(100)/O(0001) } & \makecell{4.20 \\ 3.65} & \makecell{0.69 \\ 1.57} & 1.03 \\
		Ti/Sapphire & 0.125 & 6050.00 & \makecell{Ti(0001)/Al(0001)  \\ Ti(0001)/O(0001) } & \makecell{4.25 \\ 3.70} & \makecell{0.68 \\ 1.77} & 1.10 \\
		Pb/Sapphire & 0.125 & 4578.33 & \makecell{Pb(100)/Al(0001)  \\ Pb(100)/O(0001) } & \makecell{4.15 \\ 3.60} & \makecell{0.43 \\ 0.98} & 0.49 \\
		Au/Sapphire & 0.125 & 4923.33 & \makecell{Au(100)/Al(0001)  \\ Au(100)/O(0001) } & \makecell{4.20 \\ 3.65} & \makecell{0.65 \\ 1.48} & 0.78 \\
		a-SiO$_{2}$/a-Al$_{2}$O$_{3}$ & 0.125 & 5901.67 & \makecell{Si/Al \\Si/O \\O/Al \\O/O}  & \makecell{4.20 \\ 3.65 \\3.65 \\3.10} &\makecell{6.30 \\6.20 \\8.27 \\7.61} & 6.49 \\
		Al/Diamond & 0.03 & 9320.00 & Al(100)/C(100) & 3.80 & 2.02 & 0.62 \\ 
		Ti/Diamond & 0.03 & 9203.33 & Ti(0001)/C(100) & 3.85 & 2.26 & 0.67 \\
		Pb/Diamond & 0.03 & 7731.67 & Pb(100)/C(100) & 3.75 & 1.26 & 0.32 \\ 
		Au/Diamond & 0.03 & 8076.67 & Au(100)/C(100) & 3.80 & 1.90 & 0.50 \\
		GaN/AlN & 1 & 5785.50 & \makecell{Ga(0001)/Al(0001) \\Ga(0001)/N(0001) \\N(0001)/Al(0001) \\N(0001)/N(0001)} & \makecell{4.20 \\ 3.70 \\3.70 \\3.20} & \makecell{1.46 \\ 1.14 \\1.14 \\0.85} & 7.36 \\
		TiN/MgO & 1 & 6526.83 & \makecell{Ti(111)/Mg(111) \\Ti(111)/O(111) \\N(111)/Mg(111) \\N(111)/O(111)} & \makecell{4.35 \\ 3.70 \\3.80 \\3.15} &\makecell{1.92 \\1.39 \\1.47 \\1.01}  & 10.01 \\
		\hline
	\end{tabular}
	\label{table1}
\end{table*}

\end{document}